\documentclass[aip,amsmath,amssymb,reprint]{revtex4-1}

\usepackage{graphicx}
\usepackage{dcolumn}
\usepackage{bm}

\usepackage[utf8]{inputenc}
\usepackage[T1]{fontenc}
\usepackage{mathptmx}
\usepackage{etoolbox}
\usepackage{mathrsfs}
\usepackage{soul, xcolor}
\usepackage{color, wasysym}

\usepackage{amsmath}
\usepackage{amsfonts}
\usepackage{amssymb}
\usepackage{graphicx}
\usepackage{bm}
\usepackage{textcomp}
\usepackage{srcltx}
\usepackage{amsthm} 
\usepackage{latexsym}
\usepackage{epsfig}
\usepackage{mathrsfs}

\makeatletter
\def\@email#1#2{%
 \endgroup
 \patchcmd{\titleblock@produce}
  {\frontmatter@RRAPformat}
  {\frontmatter@RRAPformat{\produce@RRAP{*#1\href{mailto:#2}{#2}}}\frontmatter@RRAPformat}
  {}{}
}%

\DeclareMathOperator{\D}{d\!}
\DeclareMathOperator{\E}{e} 
\DeclareMathOperator{\I}{i}

\newtheorem{theorem}{Theorem}
\newtheorem{definition}{Definition}

\makeatother
\begin{document}

\preprint{AIP/123-QED}

\title[Heterogeneous Cattaneo-Vernotte equation connection to the noisy voter model]{Heterogeneous Cattaneo-Vernotte equation connection to the noisy voter model}

\author{K. G\'{o}rska}
\email{katarzyna.gorska@ifj.edu.pl}

\author{A. Horzela}

\affiliation{H. Niewodnicza\'{n}ski Institute of Nuclear Physics, Polish Academy of Sciences, Division of Theoretical Physics, ul. Eliasza-Radzikowskiego 152, PL 31-342 Krak\'{o}w, Poland}

\author{D. Jankov Ma\v{s}irevi\'c}

\affiliation{School of Applied Mathematics and Informatics, University of Osijek, Trg Ljudevita Gaja 6, 31000 Osijek, Croatia}

\author{T. Pietrzak}
\affiliation{H. Niewodnicza\'{n}ski Institute of Nuclear Physics, Polish Academy of Sciences, Division of Theoretical Physics, ul. Eliasza-Radzikowskiego 152, PL 31-342 Krak\'{o}w, Poland}

\author{T. K. Pog\'any}

\affiliation{Institute of Applied Mathematics,  John von Neumann Faculty of Informatics, \'Obuda University, B\'ecsi \'ut 96/b, 1034 Budapest, Hungary}

\affiliation{Faculty of Maritime Studies, University of Rijeka Studentska 2, 51000 Rijeka, Croatia}

\author{T. Sandev}

\affiliation{Research Center for Computer Science and Information Technologies, Macedonian Academy of Sciences and Arts, Bul. Krste Misirkov 2, 1000 Skopje, Macedonia}
\affiliation{Institute of Physics, Faculty of Natural Sciences and Mathematics, Ss.~Cyril and Methodius University, Arhimedova 3, 1000 Skopje, Macedonia}
\affiliation{Department of Physics, Korea University, Seoul 02841, Republic of Korea}

\date{\today}

\begin{abstract}
We consider a heterogeneous diffusion equation and its corresponding generalization to the Cattaneo-Vernotte equation. It is derived by a combination of the continuity equation and the constitutive relation in various stochastic interpretations of the heterogeneous diffusion process. The heterogeneity in the system is introduced by considering a position-dependent diffusion coefficient. Exact results for the probability density function and the mean squared displacement are provided. The limiting case of heterogeneous diffusion is analyzed in detail, and the corresponding time-averaged mean-squared displacement is calculated. From the obtained results, an ergodicity breaking is observed. 
\end{abstract}

\maketitle

\begin{quotation}
Anomalous diffusion has been observed in various systems. It can be described by different stochastic and kinetic models in which, for example, the memory effect or the heterogeneity of the environment should be taken into consideration. We explore a model of anomalous diffusion in terms of heterogeneous diffusion and Cattaneo-Vernotte equation. The heterogeneity in the system is expressed by position-dependent diffusion coefficient. Such model may be of interest not only for description of anomalous and heterogeneous diffusion in various systems, but also for description of random processes in opinion dynamics and financial markets. 
\end{quotation}

\section{Introduction}\label{sect1}

Anomalous diffusion has been observed in various phenomena in physics, biology, chemistry, economy, etc. In biology, the motion of bacteria in heterogeneous environments, particularly photokinetic bacteria in external light fields \cite{GFrangipane18, GVizsnyiczai17} 
shows anomalous diffusive behavior as well as passive motion of lipid granules or telomeres in living cells~\cite{tabei2013intracellular} and in non-Markovian gene expressions.~\cite{vilk2024non} In chemistry with phase movement within proteins~\cite{kou2004generalized} and diffusion in lipid membranes~\cite{jeon2012anomalous} are other examples of such systems, as well as in different chemical reactions.~\cite{lindenberg2006chemical} In physics, it is related to Richardson diffusion in turbulence,~\cite{richardson1926atmospheric, ALStella23} in generalized Lotka-Voltera models of ecosystems,~\cite{GBiroli18} in charge carrier motion in amorphous semiconductors~\cite{scher1975anomalous}, diffusion in porous material,~\cite{wang2019diffusive} etc. It is also observed in different socio-economic systems, such as in opinion dynamics and financial markets modeled by the so-called voter model.~\cite{RKazakevicius21, RKazakevicius23} Anomalous diffusion in chaotic,~\cite{geisel1984anomalous} and deterministic~\cite{barkai2003aging} dynamical systems has been of interest, as well.

Anomalous diffusion can be characterized by the nonlinear temporal behavior of the mean square displacement (MSD), $\text{MSD} \propto t^{\alpha}$ ($\alpha\neq1$).~\cite{RMetzler00} The MSD can be calculated over the whole ensemble. In the (1+1)-dimensional space it reads
\begin{equation}\label{26/11/25-2}
\langle x^2(t) \rangle = \int_{-\infty}^{\infty} x^{2} p(x,t)\,dx. 
\end{equation}
Alternatively, it can also be defined by time averaging, i.e., time averaged MSD (TA-MSD), $\langle \overline{\delta^2(\mathfrak{T}, T)}\rangle = N^{-1} \sum_{j=1}^N \overline{\delta^2_j(\mathfrak{T}, T)}$, where $N$ is the number of particles in the ensemble, which in (1+1)-dimensional space $\overline{\delta^2_j(\mathfrak{T}, T)}$ is defined by~\cite{RMetzler14} \begin{equation}\label{26/11/25-3}
\overline{\delta^2_j(\mathfrak{T}, T)} = \frac{1}{T - \mathfrak{T}} \int_{0}^{T-\mathfrak{T}} [x_j(t + \mathfrak{T}) - x_j(t)]^{2} \D t. 
\end{equation}
It enables us to analyze the particle trajectory. The symbols $\mathfrak{T}$ and $T$ denote the lag time and the length of the time series $x(t)$, respectively. Furthermore, we assume that $\mathfrak{T} \ll T$. Thus, the MSDs can be calculated either by knowing the probability density function (PDF) $p(x, t)$ of the process (it describes the probability density of finding the particle at position $x\in\mathbb{R}$ in time $t \in \mathbb{R}_{+}$) or from the corresponding Langevin equation for the trajectory $x(t)$. It is well known that one can find the corresponding Fokker-Planck equation for a given Langevin equation in the presence of white noise.~\cite{risken1989fpe, vanKampen} The solutions of such Fokker-Planck equations represent the PDFs, which are non-zero at infinite distances from the initial position of the particle even for a very short time. This means that those processes have an infinite propagation speed. The finite propagation speed can be considered if one introduces the so-called telegrapher's processes.~\cite{kac1974stochastic,balakrishnan1993simple,masoliver1996finite} The standard telegrapher's process can be described by the overdamped Langevin equation with dichotomous noise from which one can derive the corresponding standard telegrapher's/Cattaneo-Vernotte (CV) equation for the PDF. Transition from a heterogeneous Langevin equation with a position-dependent diffusion coefficient and multiplicative dichotomous noise to a heterogeneous CV equation is also known for some special cases.~\cite{LAngelani19, NERatanov99}

Many studies of diffusion processes in heterogeneous, disordered, and porous media, as well as biological cells, have shown that the complex structure of the environment strongly affects particle motion, leading to anomalous dynamics. Therefore, the development of models that account for the structure of the environment is needed to describe the corresponding diffusive process. This anomalous dynamics can be the result of the memory effects of the environment,~\cite{RMetzler00} due to the presence of random potentials,~\cite{simon2013transport}
due to variations of the local diffusion coefficient in space~\cite{AGCherstvy13a} and time,~\cite{lim2002self} or due to the velocity gradient.~\cite{abdoli2024shear} 

In this paper, we consider heterogeneous diffusion and CV models with a position-dependent diffusion coefficient. Position-dependent diffusivity appears in mesoscopic approaches to transport in heterogeneous porous media,~\cite{FGjetvaj15} in relative diffusion of passive tracers in the atmosphere,~\cite{richardson1926atmospheric} in turbulent diffusion~\cite{OGBakunin04} and turbulent flows,\cite{BSawford01} etc. The finite propagation speed makes such models more realistic than the corresponding diffusion models.

The paper is organized as follows. In Section~\ref{sec20} we give some known results related to the heterogeneous diffusion equations. The corresponding Langevin and Fokker-Planck equations are presented, and the known results for the PDFs and MSDs are given. The connection of such equations to models of opinion dynamics and financial markets is also given. The Heterogeneous CV equations, as generalizations of the corresponding diffusion equations, are considered in Section~\ref{sect2}. Some known results related to PDFs and MSDs are presented and compared. In Section~\ref{sect3} we give the exact solution of the heterogeneous CV equation. The exact results for the moments and the MSD, as well as their asymptotic behaviors in the short and long time limits, are given in Section~\ref{sec_moments}. In Section~\ref{sect4} we analyze the limiting case with $\tau=0$. In Section~\ref{sect5} the TA-MSD is presented for the memory-dependent and heterogeneous diffusion equations (the case with $\tau\to 0$). The summary is provided in Section~\ref{sect6}. Some mathematical definitions, formulas, and derivations are presented in the Appendices at the end of the paper.

\section{Heterogeneous diffusion equation}\label{sec20}

\subsection{Different interpretations of heterogeneous diffusion}

The heterogeneous diffusion with a position-dependent diffusion coefficient can be described by the stochastic differential equation called the overdamped Langevin equation
\begin{equation}\label{18/01/26-2}
    \dot{x}(t) = \sqrt{2{\cal D}(x)}\,\chi(t),
\end{equation}
where ${\cal D}(x)$ is the diffusion coefficient that changes with $x$. The trajectories $x(t)$ are obtained by integrating this equation over time; however, this procedure is not unequivocal, see Refs.~\cite{JMSancho11, APacheoPozo24}. We can do it in various ways in which we may use    
\begin{align}
    \label{24/11/25-2}
    \int_t^{t + \Delta t} \sqrt{{\cal D}[x(t')]}\,\chi(t') \D t' \simeq &\sqrt{{\cal D}[(1 - \alpha)x(t) + \alpha x(t + \Delta t)]} \nonumber \\ & \times \int_t^{t + \Delta t} \chi(t') \D t',
\end{align}
where the parameter $\alpha \in [0, 1]$ specifies the stochastic interpretation of the Langevin equation.~\cite{APacheoPozo24, JMSancho11} There are widely known three types of interpretations, i.e., the so-called H\"{a}nggi-Klimontovich ({\em HK}; $\alpha = 0$),~\cite{PHanggi82, YLKlimontovich90} Stratonovich ({\em S}; $\alpha = 1/2$),~\cite{RLStratonovich66} and Ito ({\em I}; $\alpha = 1$).~\cite{KIto44} Therefore, Eq.~\eqref{18/01/26-2} can be rewritten by adding an extra term $f(\alpha; x)$, which is interpreted as a fictitious external force. Note that the extra term also appears in the It\^{o} calculus utilized to solve the stochastic differential equation. Hence, for the chosen stochastic interpretation, the Langevin equation reads
\begin{equation}\label{17/01/26-10}
    \dot{x}(t) = \sqrt{2{\cal D}(x)}\,\chi(t) + f(\alpha; x). 
\end{equation}
To find the fictitious force $f(\alpha; x)$ we associate Eq.~\eqref{17/01/26-10} to the Fokker-Planck equation in heterogeneous media,~\cite{APacheoPozo24, IMSokolov10}
\begin{equation}\label{17/01/26-6}
\partial_t W(\alpha; x,t) = \partial_x \left\{{\cal D}(x)^{1-\alpha}\, \partial_x \left[{\cal D}(x)^\alpha W(\alpha; x, t)\right] \right\}.
\end{equation}
The above equation can be written in many forms. In particular, we distinguish three of them, which are related to stochastic interpretations of $\alpha$ such that $f(\alpha; x)$ vanishes for indicated $\alpha = 0, 1/2, 1$. We call them the H\"{a}nggi-Klimontovich, Stratonovich, and It\^o forms. In the H\"{a}nggi-Klimontovich form, eq.~\eqref{17/01/26-6} reads
\begin{multline}\label{17/01/26-9}
    \partial_t W(\alpha; x,t) = \partial_x\left[{\cal D}(x)\, \partial_x W(\alpha; x, t)\right] \\ + \alpha\,\partial_x\,\left[ {\cal D }'(x) W(\alpha; x, t)\right],
\end{multline}
in the Stratonovich form it is
\begin{multline}\label{17/01/26-8}
\partial_t W(\alpha; x,t) = \partial_x\left\{\sqrt{D(x)}\, \partial_x \left[\sqrt{{\cal D}(x)} W(\alpha; x, t)\right]\right\} \\ -\big(1/2-\alpha\big) \partial_x\left[{\cal D}'(x) W(\alpha; x, t)\right],
\end{multline}
while the It\^{o} form is
\begin{multline}\label{17/01/26-7}
\partial_t W(\alpha; x,t) = \partial_x^2 \left[{\cal D}(x) W(\alpha; x, t)\right] \\ - (1 - \alpha)\, \partial_x\left[{\cal D}'(x) W(\alpha; x, t)\right],
\end{multline}
where ${\cal D}'(x) = {\rm d}{\cal D}(x)/{\rm d} x$. Thus, $f(\alpha; x) = A(\alpha) {\cal D}'(x)$ where $A(\alpha)$ equals 
\begin{multline}\label{19/01/26-1}
A_{HK}(\alpha) = -\alpha, \quad A_S(\alpha) = 1/2 - \alpha, \quad \text{and}  \\ A_I(\alpha) = 1-\alpha. 
\end{multline}
We note that the Langevin equation~\eqref{17/01/26-10} in the It\^{o} form can be integrated using the Euler–Maruyama scheme, which is commonly used for simulation of trajectories. 

Here, for the diffusion coefficient ${\cal D}(x)$, we take ${\cal D}_{0, \beta}(x) = B |x|^{2 - 2/\beta}$ with $B > 0$ and $\beta > 0$, see Refs.~\cite{NLeibovich19, AGCherstvy13, AGCherstvy13a}. The Langevin equation \eqref{17/01/26-10} for $x > 0$ reads
\begin{equation}
    \label{23/01/26-4}
    \dot{x}(t) = \sqrt{2 B} x^{1 - 1/\beta}\, \chi(t) + 2B\, A(\alpha)\, (1 - 1/\beta) x^{1 - 2/\beta}.
\end{equation}
To proceed further we rewrite Eq.~\eqref{17/01/26-10} with $f(\alpha; x) = A(\alpha) {\cal D}_{0, \beta}'(x)$ by setting $y = \int_{0}^{x}[{\cal D}_{0, \beta}(\xi)]^{-1/2}\, \D\xi$. Thus, we obtain the Langevin equation for the Bessel process
\begin{equation}\label{17/01/26-11}
    \dot{y}(t) = \sqrt{2}\,\chi(t) + \frac{2 A(\alpha) (1 - 1/\beta)}{y},    
\end{equation}
where $A(\alpha)$ depends on the stochastic interpretations parameter $\alpha$ and it is defined by Eq.~\eqref{19/01/26-1}. The fictitious external force $F(\alpha; y)$ reads $2 A(\alpha) (1 - 1/\beta) y^{-1}$ and the external potential $V(\alpha; y)$ corresponding to it is logarithmic, namely $V(\alpha; y) = 2 A(\alpha) (1/\beta - 1) \ln{y}$. The solution of the Fokker-Planck equation~\eqref{17/01/26-6} for ${\cal D}(x) \equiv {\cal D}_{0, \beta}(x)$ and $\alpha = 0, 1/2, 1$ is given by
\begin{equation} \label{19/01/26-2}
     W_{0, \beta}(\alpha; x, t) = 
     \frac{\sqrt{\pi} |x|^{2\alpha(1-\beta)/\beta}}{\Gamma(1-\nu)} \left(\frac{\beta^2}{4 B t}\right)^{1/2 - \nu} N\big(\beta |x|^{1/\beta}, t\big), 
\end{equation}
where 
\begin{equation}
    \label{31/01/26-1}
    \nu = (1-2\alpha)(1 - \beta)/2 + 1/2
\end{equation}
and $N(z, t)$ is the Gauss (normal) PDF 
\begin{equation}\label{18/01/26-1}
 N(x, \xi) = \frac{1}{\sqrt{4\pi B \xi}} \exp\Big(\!-\frac{z^{2}}{4 B \xi}\Big).
\end{equation}
For values of $\nu$ related to $\alpha = 0, 1/2, 1$ we reconstruct the three cases presented in Sections~5.1.1-5.1.3 of Ref.~\cite{TSandev22b} or for $\alpha = 1/2$ we get Eq. (2) of Ref. \cite{AGCherstvy13a}.
The corresponding MSD for $\alpha = 1/2$ shows a power-law dependence on time
\begin{equation}
    \label{19/01/26-3}
    \langle x^2(t)\rangle_{W} = \frac{\Gamma(1+2\beta)}{\beta^{2\beta}}\, \frac{(B t)^\beta}{\Gamma(1+\beta)}=\frac{\Gamma(\beta+1/2)}{\sqrt{\pi}}(2/\beta)^{2\beta}(Bt)^\beta,
\end{equation}
in which $W = W_{0, \beta}(1/2; x, t)$, see Ref. \cite{AGCherstvy13a}. It means subdiffusion for $0<\beta<1$, normal diffusion for $\beta=1$, and superdiffusion for $\beta>1$. 

We note here that the diffusion coefficient ${\cal D}_{0, \beta}(x)$ vanishes or has an infinite value  at $x=0$ depending on the value of $\beta$ chosen. To avoid this inconvenience, we generalize it to the form~\cite{NLeibovich19, AGCherstvy13} 
\begin{equation}\label{1/09/25-1}
{\cal D}_{\lambda, \beta}(x) = B (\lambda + |x|)^{2 - 2/\beta}, 
\end{equation}
where $B > 0$ is a scaling parameter of dimension $[B]={\rm length}^{2/\beta}\cdot{\rm time}^{-1}$, while $[{\cal D}_{\lambda, \beta}(x)]={\rm length}^{2}\cdot{\rm time}^{-1}$. 
The parameter $\lambda > 0$ has dimension $[\lambda]={\rm length}$ and defines the value of the cusp at $x = 0$. Specific values of $\beta$ will be discussed later. 

\subsection{Connection to voter models of opinion dynamics and financial markets}\label{sec_voter}

Various socio-economic models can be interpreted as heterogeneous diffusion phenomena.~\cite{TMLiggett13, CCastellano09} An example is the noisy voter model, used to study opinion dynamics, where an individual's opinion choice is influenced by two competing mechanisms: social imitation and spontaneous change. In social imitation (herd), a voter (agent) chooses a neighbor and adopts their opinion. This mechanism promotes order and guides the system toward a uniform state. Spontaneous opinion change imitates noise. This means that, regardless of the neighbors, a voter can spontaneously change its opinion. 

We assume that each agent is assigned a binary variable that represents its opinion, namely $0$ or $1$. Subsequently, $N$ denotes the total number of voters, $n$ is the number of agents in the state $1$ at time $t \in \mathbb{R}_+$, and thus $N-n$ is the number of agents in the state $0$. The transition rates describing the noise voter model would read~\cite{OArtime19}
\begin{align}
    \label{22/01/26-1}
    \pi^+(n) & \equiv \pi(n \to n+1) = \frac{N-n}{N} \left[\frac{A}{2} + (1-A)\frac{n}{N}\right], \nonumber \\
    \pi^-(n) & \equiv \pi(n \to n-1) = \frac{n}{N} \left[\frac{A}{2} + (1-A)\frac{N-n}{N}\right].
\end{align}
Here, $A/2$ symbolizes spontaneous opinion changes, that is, a contribution of noise. The noise update is chosen with probability $A\in[0,1]$, so the final state occurs half the time regardless of the previous state. The complementary probability $1 - A$ in the voter update describes the herd mechanism. 

In the general case, the transition probabilities $\pi^\pm(n)$ depend on specific models. It can be found by solving the corresponding Fokker-Planck equation,~\cite{AJedrzejewski19, OArtime19} which can be written in the It\^{o} form~\eqref{17/01/26-7}
\begin{equation}
    \label{22/01/26-4}
    \partial_t W(x, t) = \frac{1}{2} \partial_x^2\big[{\cal D}(x) W(x, t)\big] - \partial_x\,\big[{\cal F}(x) W(x, t)\big],
\end{equation}
where $x$ is defined through $n/N$, ${\cal F}(x) = \pi^{+}(x) - \pi^{-}(x)$ is the drift, while ${\cal D}(x) = \big[\pi^{+}(x) + \pi^{-}(x)\big]/N$ is the diffusion coefficient. For the transition rate given by Eqs.~\eqref{22/01/26-1} and $x = n/N$, one has
\begin{equation}
    \label{22/01/26-3}
    {\cal F}(x) = \frac{A}{2} (1 - 2x) \quad \text{and} \quad {\cal D}(x) = \frac{A}{2 N} + \frac{2}{N} (1-A) x (1-x).
\end{equation}
Note that Eq.~\eqref{22/01/26-3} is different from ${\cal F}(x)$ and ${\cal D}(x)$ given below eq.~(3) in Ref.~\cite{OArtime19} since we assume that the relationship between $x$ and $n/N$ is not the same as there. In the spirit of Ref.~\cite{RKazakevicius21}, we can write the corresponding Langevin equation as
\begin{equation}
    \label{23/01/26-1}
    \dot{x}(t) \simeq \left[\frac{2}{N} (1-A) x (1-x)\right]^{1/2} \chi(t) + \frac{A}{2} ( 1- 2x), 
\end{equation}
where $\chi(t)$ is a white noise. 

Generalizing the above example of the herding model,~\cite{RKazakevicius21, AKononovicius12, JRuseckas11} we can consider a two-state model obtained by a nonlinear transformation $y = [x/(1-x)]^{\beta/2} = [n/(N-n)]^{\beta/2}$ for $\beta \in \mathbb{R}$ in Eq.~\eqref{23/01/26-1}. We note that for $\beta=2$ the variable $y=x/(1-x)$ is the ratio of the number of agents in state 1 to the number of agents in state 0. Thus, eq. \eqref{23/01/26-1} can be expressed as 
\begin{equation}
    \label{23/01/26-2a}
    \dot{y}(t) = \dot{y}_+(t) + \dot{y}_-(t),
\end{equation}
where
\begin{equation}    \label{23/01/26-2}
    \dot{y}_\pm(t) = \sqrt{2 B}\, y_\pm^{1 \pm 1/|\beta|}\, \chi(t) -\frac{A |\beta|}{2}\, y_\pm^{1 \pm 2/|\beta|},
\end{equation}
$|\beta| > 0$ and the diffusion coefficient equals $B=\beta^2 (1-A)/(4 N)$. The upper sign is for $n \gg N-n$, and the lower sign for $n \ll N -n$. We point out that these equations are formally similar to eq.~\eqref{23/01/26-4}. Eqs.~\eqref{23/01/26-2} were also used in financial market modeling, where $y$ corresponds to the long–term varying component of return, $\beta<2$ means that traders hide their intentions as the market goes out of equilibrium, while $\beta>2$ means that traders hide their intentions as the market approaches equilibrium.~\cite{AKononovicius12, JRuseckas11} The Fokker-Planck equation \eqref{22/01/26-4} relevant to the Langevin equation \eqref{23/01/26-2} becomes
\begin{multline*}
    \label{23/01/26-3}
    \partial_t \widetilde{W}_\pm(y, t) = 2 B \partial_y^2\, \left[y^{2(1\pm 1/|\beta|)}\, \widetilde{W}_\pm(y, t)\right] \\ + \frac{A |\beta|}{2} \partial_y\,\left[y^{1 \pm 2/|\beta|} \widetilde{W}_\pm(y, t)\right],
\end{multline*}
see eq.~(16) in Ref.~\cite{RKazakevicius16}. The change of $y$ to $z = \beta y^{1/\beta}$, using the analogous rule as it is presented belowe eq.~\eqref{23/01/26-4}, enables us to express Eq.~\eqref{23/01/26-2} for the lower sign as   
\begin{equation}
    \label{23/01/26-5}
    \dot{z}(t) = \sqrt{2 B}\, \chi(t) + F(z), \quad F(z) = -\frac{A \beta^2}{2}\, \frac{1}{z}.
\end{equation}
Comparing it with eq.~\eqref{17/01/26-11}, one can interpret $F(z)$ as the external force for which the corresponding potential $V(z)$ ($F(z) = - V'(z)$) is $V(z) = (A\beta^2/2)\, \ln z$. Thus, we have a clear connection between the heterogeneous diffusion and the noisy voter model.

\section{Heterogeneous CV~equation}\label{sect2}

The heterogeneous CV equation can be obtained in a similar way as the heterogeneous diffusion equation from the Langevin equation 
\begin{equation}\label{24/01/26-1}
\dot{x}(t) =\upsilon(x)\,\eta(t), 
\end{equation}
where $\upsilon(x) = \sqrt{{\cal D}(x)/\tau}>0$ is the position-dependend velocity, and $\eta(t)$ is a dichotomous noise. The corresponding heterogeneous CV equation in he Stratonovich interpretation ($\alpha = 1/2$) reads~\cite{TSandev22}
\begin{align}\label{26/11/25-9}
\tau\, \partial_t^2& P(1/2; x, t) + \partial_t P(1/2; x, t) \nonumber\\
&= \partial_x \left\{\sqrt{{\cal D}_{0, \beta}(x)}\; \partial_x\,\Big[\sqrt{{\cal D}_{0, \beta}(x)} P(1/2; x, t) \Big] \right\},
\end{align}
where ${\cal D}_{0, \beta}(x)$ is the position-dependent diffusion coefficient given by Eq.~\eqref{1/09/25-1} for $\lambda = 0$, and $\tau$ denotes the time lag. Its solution can be found in Eq.~(29) of Ref.~\cite{TSandev22} and its generalized version in Eq.~(9) of Ref.~\cite{LAngelani19}, 
\begin{align}\label{23/04/25-1}
   P(1/2;\, & x, t) = \frac{|x|^{1/\beta - 1}}{2} \E^{-t/(2\tau)} \Big\{\delta\left(\upsilon t - \beta |x|^{1/\beta}\right) + \frac{1}{2\tau\upsilon} \nonumber \\
   & \times \Theta\Big(\!t - \frac{\beta |x|^{1/\beta}}{\upsilon}\!\Big) \Big[I_{0}\Big(\!\frac{\Lambda_{\beta}}{2\tau\upsilon}\!\Big) + \frac{\upsilon t}{\Lambda_{\beta}} I_{1}\Big(\!\frac{\Lambda_{\beta}}{2\tau\upsilon}\!\Big)\!\Big]\!\Big\},
  \end{align}
where 
\begin{equation}\label{4/09/25-1}
	\Lambda_\beta = \sqrt{\upsilon^{2} t^2 - \beta^{2} |x|^{2/\beta}},
\end{equation}
and $\upsilon = \sqrt{B/\tau}$. It is non-zero only in a finite region $\Delta_{\beta}(t) =  x \in \left(- (\upsilon t/\beta)^{\beta}, (\upsilon t/\beta)^{\beta}\right)$, which is a consequence of the finite propagation speed encoded in the CV equation. Therefore, this is a more realistic model than the corresponding diffusion model. Note that for $\beta = 1$, the solution reduces to the solution $P_{\rm CV}(x, t)$ of the CV equation. 

The MSD calculated for $P(1/2; x, t)$ is proportional to the three-parameter Mittag-Leffler function $E_{a, b}^c(\cdot)$ (see Appendix~\ref{Ap1}), and is given by~\cite{TSandev22}
\begin{equation}\label{17/01/26-12}
   \langle x^2(t) \rangle_{P} = \left(\frac{\upsilon}{\beta}\right)^{2\beta} \Gamma(1+2\beta)\, t^{2\beta}\, E_{1, 1+2\beta}^{\beta}(-t/\tau).
\end{equation}
Asymptotically, similarly to the memory-dependent case, MSD behaves as
\begin{align}\label{18/01/26-5}
    \langle x^{2}(t)\rangle_{P}\sim\left\lbrace
    \begin{array}{ll}
    t^{2\beta},     & \text{for $t\ll\tau$}, \\
    t^\beta     & \text{for $t\gg\tau$}.
    \end{array}\right.
\end{align}

As we mentioned, there is a Langevin description of the process governed by the CV equation~\eqref{26/11/25-9} as a generalization of the heterogeneous diffusion equation in the Stratonovich interpretation.~\cite{LAngelani19} However, Eq.~\eqref{24/01/26-1} can not be used for generalization of the heterogeneous diffusion equation in the cases where $\alpha \neq 1/2$. For these cases, the finite propagation speed will be introduced through the continuous equation and constitutive relation. Namely, following by Cattaneo and Vernotte agruments, see Refs. \cite{CRCattaneo48, CRCattaneo58, PVernotte58}, we introduce
\begin{equation}
    \label{1/01/26-1}
    j(x, t) + \tau\, \partial_t j(x, t) = -{\cal D}(x)^{1-\alpha} \partial_x\left[{\cal D}(x)^{\alpha} P(\alpha; x, t)\right], 
\end{equation}
where its left-hand side can be treated as the first two terms in the Taylor expansion of the probability flux $j(x,t+\tau)$ delayed by the lag time $\tau$, i.e.,  $j(x,t+\tau) \sim j(x,t)+\tau\,\partial_t j(x,t)$. Next, using the continuity equation $\partial_t P(\alpha;x,t) = -\partial_x j(x,t)$, we arrive to the following heterogeneous CV equation
\begin{align}\label{17/12/24-1}
\tau\, \partial_t^2 P(\alpha; x, t) & + \partial_t P(\alpha; x, t) \nonumber\\ &= \partial_x \left\{{\cal D}(x)^{1-\alpha} \partial_x\big[{\cal D}(x)^{\alpha} P(\alpha; x, t)\big] \right\}. 
\end{align}
Since we already discussed the case with $\alpha=1/2$, we can rewrite the right-hand side (RHS) of the constitutive relation in the form
\begin{multline}
    \label{18/01/26-4}
    \text{RHS of Eq.~\eqref{1/01/26-1}} = -\sqrt{{\cal D}(x)} \partial_x \left[\sqrt{{\cal D}(x)}\;  P(\alpha; x, t)\right]\\ 
    + (1/2 - \alpha) {\cal D}'(x) P(\alpha; x, t),
\end{multline}
compare with Section~II of Ref.~\cite{LAngelani19}. In the It\^o form the RHS of the constitutive relation would read
\begin{multline}
    \label{26/01/26-5}
    \text{RHS of Eq.~\eqref{1/01/26-1}} = - \partial_x[{\cal D}(x) P(\alpha; x, t)] \\ + (1-\alpha) {\cal D}'(x) P(\alpha; x, t).
\end{multline}
In both cases, the term containing position-dependent diffusion coefficient ${\cal D}(x)$ is supplemented by the position-dependent fictitious term $f(x)$ defined below Eq.~\eqref{17/01/26-7} which with respect to the sign of ${\cal D}'(x)$ acts against or accelerates the diffusion caused by the diffusion term. This term vanishes in Eq.~\eqref{18/01/26-4} for $\alpha = 1/2$ and in Eq.~\eqref{26/01/26-5} for $\alpha = 1$ or in both equations for the constant diffusion coefficient ${\cal D}_{\lambda, \beta}(x) = B$, which is obtained for $\beta = 1$.

\section{Solution of heterogeneous CV equation}\label{sect3}

In this section, we solve eq.~\eqref{17/12/24-1} for the diffusion coefficient ${\cal D}(x) = {\cal D}_{\lambda, \beta}(x)$ given by Eq.~\eqref{1/09/25-1}, which is 
equipped with the fundamental initial conditions ($P_{\lambda, \beta}(\alpha; x, t=0) = \delta(x)$ and $\left.\dot{P}_{\lambda, \beta}(\alpha; x, t)\right|_{t=0} = 0$), and for boundary conditions stating that $P_{\lambda, \beta}(\alpha; x, t)$ vanishes at $\pm\infty$. The boundary conditions allow us to present the solution of Eq.~\eqref{17/12/24-1} in the form
\begin{align}\label{6/04/25-1}
    \widehat{P}_{\lambda, \beta}&(\!\alpha; x, s)  = \frac{\lambda^{(\nu-1)/\beta}}{2 \upsilon} (\lambda + |x|)^{a} 
    \nonumber\\  &\times \frac{s + 1/\tau}{\sqrt{s^2 + s/\tau}}  \frac{K_{\nu}\left( \frac{\beta}{\upsilon} (\lambda + |x|)^{1/\beta} \sqrt{s^2 + s/\tau}\right)}{K_{\nu-1}\left(\frac{\beta}{\upsilon} \lambda^{1/\beta} \sqrt{s^2 + s/\tau}\right)},
\end{align}
derived in Appendix \ref{Ap2}. The function $P(\alpha; x, t)$ has the indices $\lambda$ and $\beta$ to indicate its dependence on these parameters. Here, $K_{\mu}(\cdot)$ denotes the modified Bessel function of the second kind, also known as the Macdonald function, while the parameters $a$ and $\nu$ are given by
\begin{align}\label{5/09/25-1}
	a = 1/2 + (1 + \alpha)&(1 - \beta)/\beta, 
\end{align}
and eq. \eqref{31/01/26-1}, espectively.

The function $P_{\lambda, \beta}(\alpha; x, t) = \mathscr{L}^{-1}[\widehat{P}_{\lambda, \beta}(\alpha; x, s); t]$ represents a PDF since it is normalized and nonnegative. Normalization of $P_{\lambda, \beta}(\alpha; x, t)$ can be shown by making direct calculations, which are presented in Appendix \ref{Ap3}. The proof of its nonnegativity is a challenging task. It can be demonstrated using the Bernstein theorem (see Appendices \ref{Ap1} and \ref{Ap4}), which uniquely combines a nonnegative function with a completely monotonic function (CMF) via the Laplace transform. For instance, let us represent Eq.~\eqref{6/04/25-1} for $\tau \ll 1$ in the form
\begin{equation}\label{14/10/25-1}
P_{\lambda, \beta}(\alpha; x, s)\; {\underset{{\tau\to 0}}{\longrightarrow}}\; \frac{1}{\sqrt{s}} \frac{K_{\nu}(\sigma_{\lambda, \beta}(x) \sqrt{s})}{K_{\nu - 1}(\sigma_{\lambda, \beta}(0) \sqrt{s})},
\end{equation}
in which $\nu > 0$ and $\sigma_{\lambda, \beta}(x)$ denotes the positive increasing function $(\beta/\upsilon) (\lambda + |x|)^{1/\beta}$. Thus, it implies that $\sigma_{\lambda, \beta}(x) \geq \sigma_{\lambda, \beta}(0)$.  Then, rewriting it as
\begin{equation}\label{13/10/25-1}
	\left(\frac{1}{\sqrt{s}} \frac{K_{\nu}(\sigma_{\lambda, \beta}(x) \sqrt{s})}{K_{\nu-1}(\sigma_{\lambda, \beta}(x) \sqrt{s})}\right)\, \left(\frac{K_{\nu-1}(\sigma_{\lambda, \beta}(x) \sqrt{s})}{K_{\nu-1}(\sigma_{\lambda, \beta}(0) \sqrt{s})}\right)
\end{equation}
and employing Theorems 9 and 10 in Ref. \cite{KSMiller01} or Refs. \cite{MEHIsmail76, MEHIsmail77} we show that the function in Eq.~\eqref{13/10/25-1} is a CMF as a product of two CMFs. The function $\widehat{P}_{\lambda, \beta}(\alpha; x, s)$ in the limit of $\tau \to \infty$ can be expressed as 
\begin{equation}\label{14/10/25-2}
\widehat{P}_{\lambda, \beta}(\alpha; x, s)\; {\underset{{\tau\to\infty}}{\longrightarrow}}\; \frac{K_{\nu}(\sigma_{\lambda, \beta}(x) s)}{K_{\nu - 1}(\sigma_{\lambda, \beta}(0) s)}, 
\end{equation}
which is a CMF for $\nu\in(0, 3/2]$, see Appendix~\ref{Ap4}. From here, it can be shown that Eq.~\eqref{14/10/25-2}, in which $s$ is replaced by another Bernstein function $(s^{2} + s/\tau)^{1/2}$, is also a CMF for $\nu\in(0, 3/2]$. The proof of the Bernstein character of $(s^{2} + s/\tau)^{1/2}$ can be found in Section~4~A of Ref.~\cite{KGorska20}. To show that Eq.~\eqref{6/04/25-1} is a CMF we should also proved that
\begin{equation}\label{15/10/25-1}
\frac{s+1/\tau}{\sqrt{s^{2} + s/\tau}} = \left[1 + (\tau s)^{-1}\right]^{1/2}
\end{equation}
is a CMF; that results from Eq.~(1.19) of Ref.~\cite{KSMiller01}. From the fact that the product of CMFs is a CMF appears that $\widehat{P}_{\lambda, \beta}(\alpha; x, s)$ is a CMF for $\nu\in(0, 3/2]$, as well. Hence, its inverse Laplace transform is nonnegative.

We note that we are able to make the analytical inversion of the Laplace transform of Eq.~\eqref{6/04/25-1} only in two cases: (i) either $\lambda=0$, or (ii) $\lambda \neq 0$ while $\nu$ is a natural multiple of $1/2$. 

\subsection{Solution for $\lambda = 0$}\label{sect3.1}

For $\beta > 0$, we utilize the property of the modified Bessel function of the second kind, which states that $K_{\mu}(\cdot) = K_{-\mu}(\cdot)$, see Ref. \cite{ABaricz17}. Applying the asymptotic behavior of $K_\mu(x)$ at $x=0$ given by Eq.~\eqref{7/04/25-2}, we have 
\begin{align}
    \label{9/09/25-1}
    \widehat{P}_{0, \beta}(&\alpha; x, s) = \left(\frac{\beta}{2 \upsilon}\right)^{1-\nu}\, \frac{|x|^{a}}{\upsilon\, \Gamma(1-\nu)} \nonumber\\ &\times \frac{s + 1/\tau}{(s^2 + s/\tau)^{\nu/2}} K_\nu\left(\frac{\beta |x|^{1/\beta}}{\upsilon}\sqrt{s^2 + s/\tau}\right),
\end{align}
which is valid for $1-\nu > 0$, i.e., $\nu < 1$. This gives a restriction on the parameter $\beta$ such that $(1 - 2\alpha)(1 - \beta) < 1$. Thus, $\alpha = 0$ implies $\beta >0$; while for $\alpha = 1/2$ we have identity, and thus $\beta > 0$; finally, $\alpha = 1$ yields $\beta \in(0, 2)$. The inverse Laplace transform of Eq.~\eqref{9/09/25-1}, obtained in Appendix~\ref{Ap5}, reads
\begin{align}\label{10/09/25-1}
P_{0, \beta}&(\alpha; x, t)  = \Big(\frac{4\tau \upsilon^2}{\beta^2}\Big)^{\nu-1/2} \frac{\sqrt{\pi}\,|x|^{2\alpha(1/\beta-1)}}{2\Gamma(1-\nu)} \E^{-t/(2\tau)} \nonumber \\ & \times \left\{ \delta\left(\upsilon t - \beta |x|^{1/\beta}\right) + \frac{(\Lambda_{\beta}/\upsilon)^{\nu - 1/2}}{2\tau\upsilon} \Theta\left(t - \beta |x|^{1/\beta}/\upsilon\right) \right. \nonumber \\
& \left.\times \left[I_{\nu-1/2}\left(\frac{\Lambda_{\beta}}{2\tau \upsilon}\right) + \frac{\upsilon t}{\Lambda_{\beta}} I_{\nu - 3/2}\left(\frac{\Lambda_{\beta}}{2\tau \upsilon}\right)\right]\right\},
\end{align}
where $\Lambda_\beta$ is defined by Eq.~\eqref{4/09/25-1}. Taking into account $I_{-n}(\cdot) = I_n(\cdot)$, where $n =0, 1, 2, \ldots$, Eq.~\eqref{10/09/25-1} for $\nu = 1/2$ is given by Eq.~\eqref{23/04/25-1} and describes the telegraphers process.~\cite{EOrsinger04, MDOvidio18} Furthermore, $P_{0, \beta}(\alpha; x, t)$ is a nonnegative function that is non-zero inside $\Delta_{\beta}(t)$ defined below Eq.~\eqref{4/09/25-1}, and vanishes outside of this region. 

\begin{figure}
\includegraphics[scale=0.44]{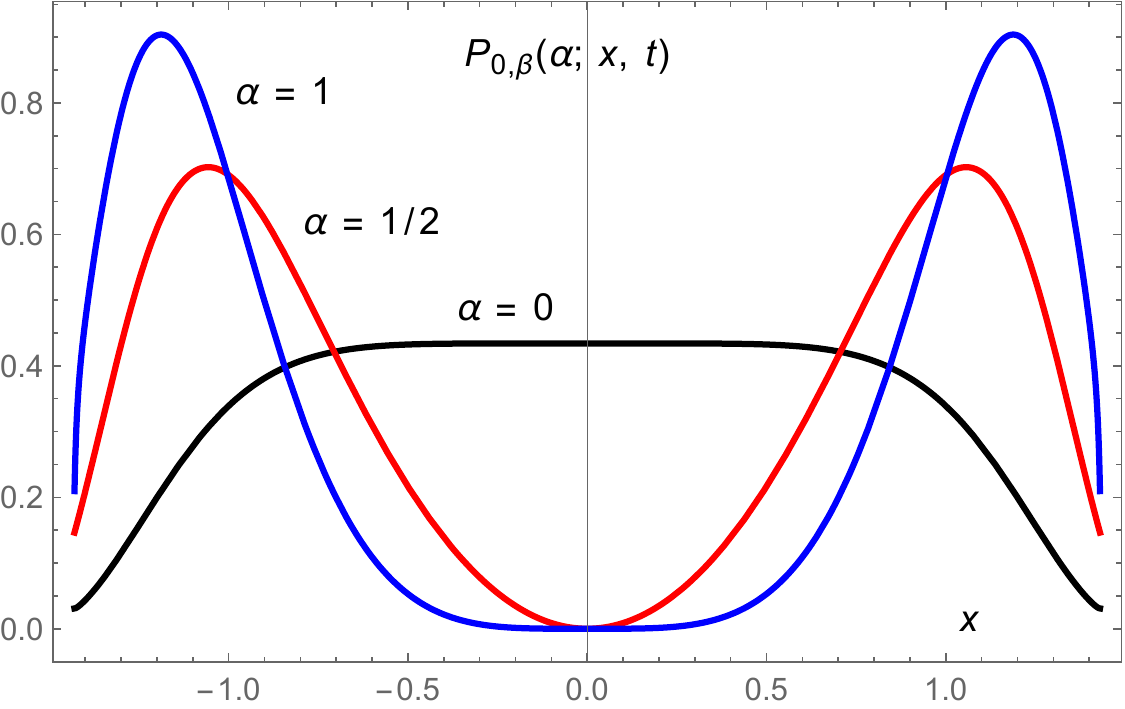}
\caption{\label{fig1} PDF $P_{0, \beta}(\alpha; x, t)$ for $\beta = 1/3$, $\tau = 0.1$, $v = 1$, $t=1$, and different values of $\alpha$: $\alpha=0$ (black curve), $\alpha = 1$ (blue curve) and $\alpha = 1/2$ (red curve).}
\end{figure}

\begin{figure}
\includegraphics[scale=0.41]{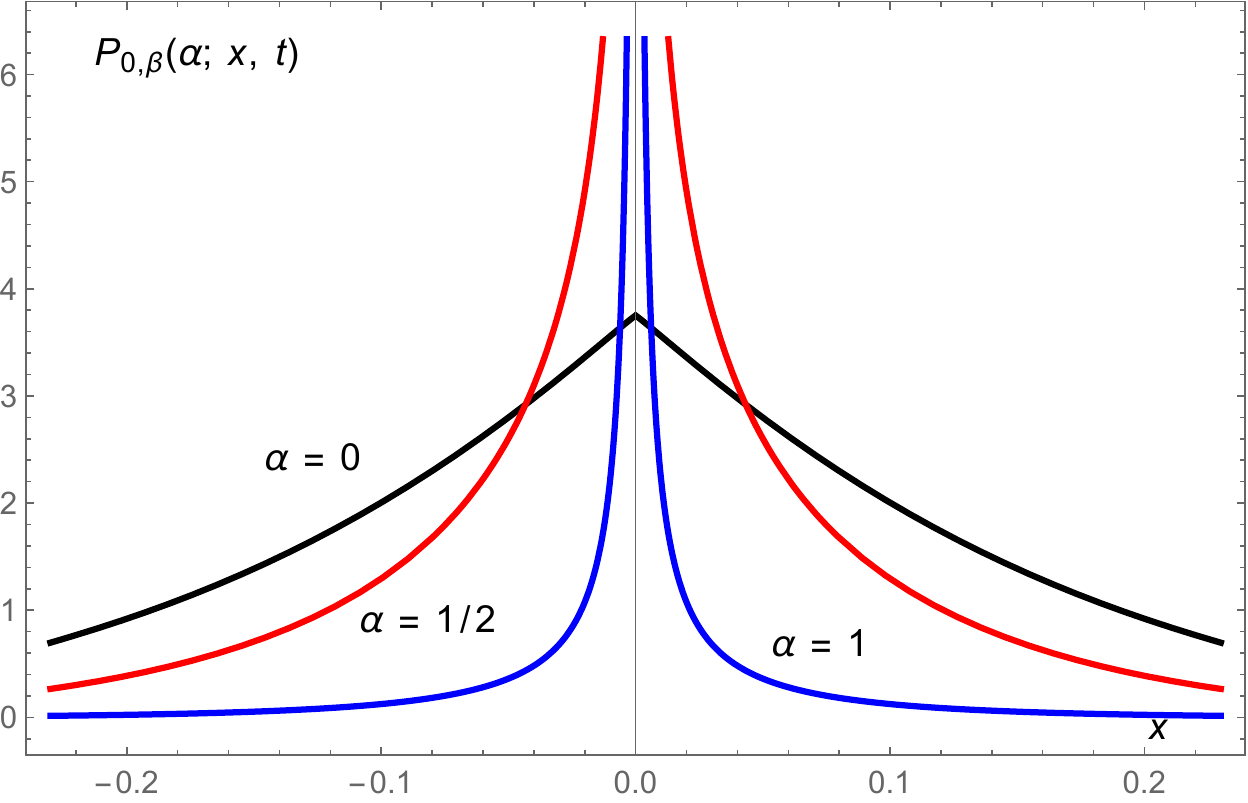}
\caption{\label{fig2} PDF $P_{0, \beta}(\alpha; x, t)$ for $\beta = 1.9$, $\tau = 0.1$, $v = 1$, $t=1$, and different values of $\alpha$: $\alpha=0$ (black curve), $\alpha = 1$ (blue curve) and $\alpha = 1/2$ (red curve).}
\end{figure}

Graphical representations of $P_{0, \beta}(\alpha; x, t)$ in various stochastic interpretations, for $\beta = 1/3$ and $\beta = 1.9$ are given in Figs.~\ref{fig1} and~\ref{fig2}, respectively. Here, the diffusion coefficient ${\cal D}_{0, 1/3}(x)$ is proportional to $|x|^{-4}$ and for $\beta = 1.9$ we have ${\cal D}_{0, 1.9}(x) \propto |x|^{0.95}$.
Here, we can distinguish two different behaviors for fixed $\alpha\neq 0$. Namely, for $\beta\in (0, 1)$ we observe two peaks symmetric with respect to $x = 0$. For $\beta > 1$ (note that for $\alpha = 1$, the parameter $\beta$ is only in the range $(0, 1)$) we have the symmetric behavior presented in Fig.~\ref{fig2}, with cusp at $x=0$ and approaching infinity as $|x|$ decreases. As shown in Figs.~\ref{fig1} and~\ref{fig2}, completely different behaviors for the case with $\alpha=0$ is observed. 

\subsection{Solution for $\lambda > 0$ and $\nu = 1/2$ and $\nu = 3/2$ }\label{sect3.2}

\noindent
{\bf (i)} For $\nu = 1/2$, that is the telegraphers process, we use Eq.~\eqref{7/04/25-1} which applied to eq.~\eqref{6/04/25-1} yields
\begin{equation}\label{12/09/25-2}
\widehat{P}_{\lambda, \beta}(\alpha = 1/2; x, s) = (\lambda + |x|)^{\frac{1}{\beta} - 1} \widehat{P}_{\rm CV}(y, s),
\end{equation}
where 
\begin{equation}\label{4/09/25-2}
y = \beta \big[(\lambda + |x|)^{1/\beta} - \lambda^{1/\beta}\big]
\end{equation}
and 
\begin{equation}\label{12/09/25-1}
\widehat{P}_{\rm CV}(y, s) = \frac{1}{2\upsilon} \frac{s + 1/\tau}{\sqrt{s^{2} + s/\tau}} \exp\left(\!-\frac{y}{\upsilon} \sqrt{s^{2} + s/\tau}\right).
\end{equation}
The inverse Laplace transform of eq.~\eqref{12/09/25-2} gives
\begin{equation}\label{12/09/25-4}
P_{\lambda, \beta}(1/2; x, t) = (\lambda + |x|)^{1/\beta - 1} P_{\rm CV}(y, t),
\end{equation} 
where $P_{\rm CV}(y, t) = P_{0, 1}(1/2; y, t)$. The solution \eqref{12/09/25-4} can also be found in Ref.~\cite{LAngelani19}, see Eq.~(9) for $\upsilon(x) = [{\cal D}_{\lambda, \beta}(x)/\tau]^{1/2}$. It has a non-zero value for $|y| < \upsilon t$ which implies $x\in\Delta_{\lambda, \beta}$, where  $$\Delta_{\lambda, \beta} = \left(-(\upsilon t/\beta + \lambda^{1/\beta})^{\beta} + \lambda,\; (\upsilon t/\beta + \lambda^{1/\beta})^{\beta} - \lambda\right).$$ 
\begin{figure}
    \centering
    \includegraphics[scale=0.45]{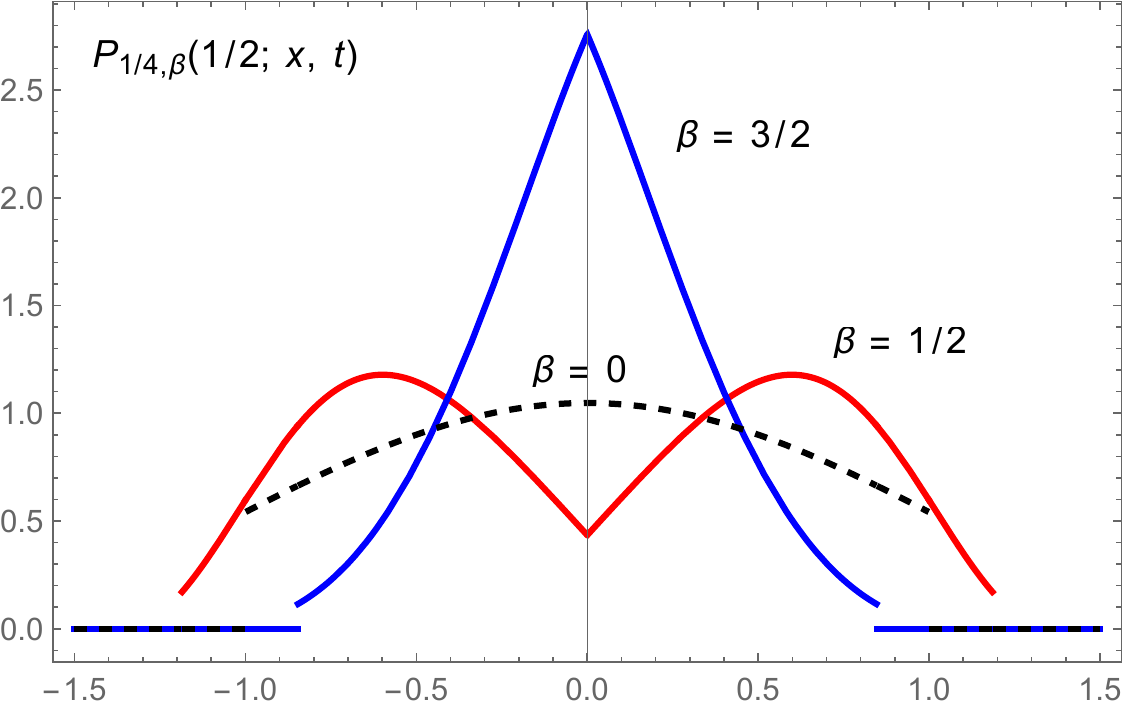}
    \caption{Plot of $P_{\rm CV}(x, t) = P_{0, 1}(1/2; x, t)$ (the black dashed curve), $P_{1/4, 1/2}(1/2; x, t)$ (the red solid curve), and $P_{1/4, 3/2}(1/2; x, t)$. PDFs are plotted for $\tau = 0.1$, $\upsilon = 1$, and $t =1$.}
    \label{fig3}
\end{figure}

Graphical representation of the PDFs $P_{\rm CV}(x, t) = P_{0, 1}(1/2; x, t)$, that solves the CV equation, and $P_{\lambda, \beta}(1/2; x, t)$ (Eq.~\eqref{12/09/25-4}) being the solution of heterogeneous CV equation is given in Fig.~\ref{fig3}. We emphasise that PDFs $P_{\lambda, \beta}(1/2; x, t)$ for $t = 1$ and $\beta = 1/2$ as well as $t = 1$ and $\beta = 3/2$ contain wave fronts, i.e., that is, they are nonzero only in finite region $\Delta_{\lambda, \beta}$, which for $\beta = 1/2$ is $(-1.19, 1.19)$ and $\beta = 3/2$ is equal to $(-0.85, 0.85)$.  

\smallskip

\noindent 
{\bf (ii)} The case with $\nu = 3/2$, where $\nu$ is defined by Eq.~\eqref{31/01/26-1}, can be obtained $\alpha = 1$ and $\beta = 3$. However, taking into account the property of the Macdonald function, we should also consider the case $\nu = -3/2$, which is realized for $\alpha = 0$ and $\beta = 5$. Thus, Eq.~\eqref{7/04/25-1} employed to Eq.~\eqref{6/04/25-1} leads to
\begin{align}\label{12/09/25-5}
\widehat{P}_{\lambda, 3}&(1; x, s)  \equiv\, \widehat{P}_{\lambda, 5}(0; x, s) = \frac{\lambda^{1/\beta}}{(\lambda + |x|)^{1/(2\beta) - a}} \widehat{P}_{\rm CV}(y, s)  \nonumber\\
&+ \frac{\lambda^{1/\beta}}{2 \beta (\lambda + |x|)^{3/(2\beta) - a}}  s^{-1}  \exp\Big(\!-\frac{y}{\upsilon} \sqrt{s^{2} + s/\tau}\Big).
\end{align}
Its inverse Laplace transform, by using Eq.~\eqref{12/09/25-6}, yields
\begin{align}
    \label{12/09/25-10}
 P_{\lambda, 3}&(1; x, t)  \equiv P_{\lambda, 5}(0; x, t) = \frac{\lambda^{1/\beta}}{(\lambda + |x|)^{1/(2\beta) - a}} P_{\rm CV}(y, t) \nonumber \\
 & + \frac{\lambda^{1/\beta}}{2 \beta (\lambda + |x|)^{3/(2\beta) - a}} \left\{\E^{- y/(2\tau\upsilon)} \Theta(t - y/\upsilon) \right.\nonumber \\
 &+ \left.\frac{y}{2\tau}   \int_{y/\upsilon}^t \frac{\E^{-\xi/(2\tau)}}{\sqrt{\upsilon^2 \xi^2 - y^2}} I_1\left(\frac{\sqrt{\upsilon^2 \xi^2 - y^2}}{2\tau\upsilon} \right) d\xi\right\}. 
\end{align}

\subsection{Moments and MSD}\label{sec_moments}

Since $P_{\lambda, \beta}(\alpha; x, t)$ is a symmetric function with respect to $x = 0$, all odd moments are equal to zero, while the even moments $\langle x^{2q}(t)\rangle = \int_\mathbb{R} x^{2m} p(x, t) \D x$ for $m = 0, 1, \ldots$ are non-vanishing.

The even moments for $P_{\lambda, \beta}(\alpha; x, t)$, in which $\lambda = 0$, are given by 
\begin{align}
    \label{10/09/25-3}
    \langle x^{2m}(t) \rangle  & = \mathscr{L}^{-1}\Big[2 \int_{0}^{\infty} x^{2m}\, \widehat{P}_{0, \beta}(\alpha; x, s) \D x; t \Big]\nonumber \\ 
    & = \Big(\frac{2 \upsilon}{\beta}\Big)^{\!2\beta m}\, \frac{\Gamma(1 + \beta m) \Gamma(1 + \beta m - \nu)}{\Gamma(1-\nu)} \nonumber \\ & \times t^{2\beta m} E^{\beta m}_{1, 1 + 2\beta m}(-t/\tau).
\end{align}
For $\alpha = 1/2$ and $m = 1$ it reduces to $\langle x^2(t)\rangle_{P}$ given by Eq.~\eqref{17/01/26-12}. The Prabhakar function $t^{b-1} E^{c}_{a, b}(x)$ is related to the three-parameter Mittag-Leffler function $E_{a, b}^c(\cdot)$, see Appendix~\ref{Ap1}. Note that $\alpha$ is hidden in the parameter $\nu$ and appears in the coefficient before the Prabhakar function. Therefore, it does not affect the time behavior of $\langle x^{2m}(t) \rangle$. The MSD is obtained from Eq.~\eqref{10/09/25-3} for $m = 1$ and it reads
\begin{align}
    \label{10/09/25-3msd}
    \langle x^{2}(t) \rangle  & = \left(\frac{2 \upsilon}{\beta}\right)^{\!2\beta }\, \frac{\Gamma(1 + \beta) \Gamma(1 + \beta - \nu)}{\Gamma(1-\nu)} \nonumber \\ & \times t^{2\beta} E^{\beta}_{1, 1 + 2\beta}(-t/\tau).
\end{align}
The asymptotic behavior of the moments is $\langle x^{2m}(t) \rangle\propto t^{2\beta m}$ for $t \ll \tau$ and $\langle x^{2m}(t) \rangle \propto t^{\beta m}$ for $t \gg \tau$. Thus, the corresponding MSD behaves as $\langle x^{2}(t) \rangle\propto t^{2\beta}$ for short times and as $\langle x^{2}(t) \rangle\propto t^{\beta}$ for long times, thus it has a characteristic anomalous crossover dynamics. 

To complete the results, we calculated the MSD for $P_{\lambda, \beta}(1/2; x, t)$ given by Eq.~\eqref{12/09/25-4} for $\beta = 1/2$ and $\beta = 3/2$. They read
\begin{align}
    \label{25/01/26-1}
    \langle x^2(t)\rangle_{P_{\lambda, 1/2}} & = 
 2\, \upsilon\, t E_{1, 2}^{1/2}(-t/\tau)  \nonumber \\ & + 2 \lambda^2 - 2 \lambda \int_0^\infty \sqrt{2y + \lambda^2} P_{\rm CV}(y, t)\, \D y
\end{align}
and
\begin{align}
    \label{26/01/26-1}
     \langle x^2(t)\rangle_{P_{\lambda, 3/2}} & = \frac{16}{9} (\upsilon t)^3 E_{1, 4}^{3/2}(-t/\tau) + \frac{8\lambda^{2/3}}{3} (\upsilon t)^2 E_{1, 3}^1(-t/\tau) \nonumber \\ & 
     + 2 \lambda^{4/3}\, t E_{1, 2}^{1/2}(-t/\tau) + 2 \lambda^2 \nonumber \\ &  - 2 \lambda \int_0^\infty \Big(\!\frac{2 y}{3} + \lambda^{2/3}\Big)^{3/2} P_{\rm CV}(y, t)\, \D y.
\end{align}
For $\lambda = 0$ they reduce to Eq.~\eqref{10/09/25-3msd} for $\beta = 1/2, 3/2$. 

We also note that the long time limit $t\gg\tau$ of the MSD \eqref{10/09/25-3msd} (also for the moments \eqref{10/09/25-3}) should reproduced to the case with $\tau=0$, i.e., the heterogeneous diffusion process,
\begin{align}
    \label{10/09/25-3-2-2}
    \langle x^{2}(t) \rangle = \left(2/\beta\right)^{\!2\beta}\, \frac{\Gamma(1 + \beta - \nu)}{\Gamma(1-\nu)} (B\, t)^{\beta}
\end{align}
and for $\alpha = \nu = 1/2$ it reduces to $\langle x^2(t)\rangle_W$ given by Eq.~\eqref{19/01/26-3}.

The long time limit for the heterogeneous CV equations yields the same behavior for the MSD as that for the heterogeneous diffusion equations. So, in the next section we will study the limiting behavior of the heterogeneous diffusion equation for $\alpha = 0, 1/2, 1$.

\subsection{Limit approach to the heterogeneous diffusion equation}\label{sect4}

In Ref.~\cite{KGorska21}, it is shown that the solution of the homogeneous CV equation tends  in the limit of $\tau \ll 1$ to the solution of the anomalous diffusion. The same effect is observed for $|x| \ll \upsilon t$, see Ref. \cite{HDWeymann67}. Therefore, in the case of the heterogeneous CV equation, we should meet a similar behavior. 

For $\lambda = 0$ the term containg the $\delta$-Dirac distribution in Eq.~\eqref{10/09/25-1} can be omitted for $x\in \Delta_\beta(t)$. Then, we utilize the asymptotic behavior at infinity of the modified Bessel function of the first kind, which is
\begin{multline}
    \label{2/11/25-1}
    I_\mu\left(\frac{1}{2 \tau} \sqrt{t^2 - \beta^2 |x|^{2\beta}/\upsilon^2}\right)  \stackrel{{\scriptscriptstyle{\tau\to 0}}}{\approx} \sqrt{\frac{\tau}{\pi t}} \left(1 - \frac{\beta^2 |x|^{2\beta}}{\upsilon^2 t^2}\right)^{\pm \mu/2}  \\ 
    \times \exp\left(-\frac{\beta^2 |x|^{2\beta}}{4 t \upsilon^2 \tau} + \frac{t}{2\tau}\right),
\end{multline}
where $\mu\in\mathbb{R}$ and $\alpha = 0, 1/2, 1$. Performing the direct calculations in which $\sigma_{\lambda, \beta}(\cdot)$ is defined below Eq.~\eqref{14/10/25-1}, it can be shown that 
\begin{equation}\label{2/11/25-2}
    P_{0, \beta}(\alpha; x, t)\, \underset{{\scriptscriptstyle{\tau\to 0}}}{\longrightarrow}\, W_{0, \beta}(\alpha; x, t) 
\end{equation}
where $W_{0, \beta}(\alpha; x, t)$ for $\alpha = 0, 1/2, 1$ is defined by eq. \eqref{19/01/26-2}.

Moreover, using the results of Ref.~\cite[Section~III~A]{KGorska21} it can be shown that $P_{\lambda, \beta}(1/2; x, t) $ approaches to $(\lambda + |x|)^{1/\beta - 1} N(y, t)$ for $\tau\to 0$, where $y$ is defined by Eq.~\eqref{4/09/25-2}. In the case of Eq.~\eqref{12/09/25-10}, we have
\begin{align}\label{5/11/25-1}
	P_{\lambda, 3}&(1; x, t)  \equiv \, P_{\lambda, 5}(0; x, t)
    \, \underset{{\scriptscriptstyle{\tau\to 0}}}{\longrightarrow}\,  \frac{\lambda^{1/\beta}}{(\lambda + |x|)^{1/(2\beta) - a}} N(y, t) \nonumber \\
	& + \frac{\lambda^{1/\beta}}{2 \beta (\lambda + |x|)^{3/(2\beta) - a}} \left[1 + {\rm erfc}\left(\frac{y}{2\sqrt{B t}}\right)\right].
\end{align}
Generally, the transformation of $P_{\lambda, \beta}(\alpha; x, t)$ to the solution of the heterogeneous diffusion equation for $\lambda \neq 0$ can only be easily seen in the Laplace space. Then, in Eq.~\eqref{6/04/25-1} the term $s/\tau$ dominates $s^2$ such that the latter one can be omitted. Note that here $\upsilon^2\tau=B$. The solution $\widehat{P}_{\lambda, \beta}(\alpha; x, s)$ for $\tau\to0$, denoted as $\widehat{W}_{\lambda, \beta}(\alpha; x, s)$ and satisfying Eq.~\eqref{17/01/26-7} for ${\cal D}_{\lambda, \beta}(x)$ defined by Eq.~\eqref{1/09/25-1}, is the Stieltjes function being the subclass of the CMFs.~\cite{MEHIsmail76, MEHIsmail77} Its inverse Laplace transform is nonnegative. Its normalization is preserved, i.e., $\mathscr{L}^{-1}[\int_{-\infty}^{\infty} \widehat{W}_{\lambda, \beta}(\alpha; x, s) dx; t] = 1$, see Appendix~\ref{Ap3}. Thus, $W_{\lambda, \beta}(\alpha; x, t) = \mathscr{L}^{-1}[\widehat{W}_{\lambda, \beta}(\alpha; x, s); t]$ in the limit of $\tau\to 0$ is a PDF. 

\subsection{TA-MSD}\label{sect5}

Since the PDF $W_{0, \beta}(\alpha; x, t)$ is a symmetric function, all its odd moments are equal to zero. It means that both the mean value and the time-average of $x(t)$ taken on the ensemble vanish. Thus, in the variance $\varsigma^2 = \langle x^2(t) \rangle - \langle x(t) \rangle^2$ and $\overline{\varsigma^2} = \overline{\langle x^2(t) \rangle} - \overline{\langle x(t) \rangle^2}$ the MSD and TA-MSD are important. We calculate the MSD using Eq.~\eqref {10/09/25-3-2-2}; in the following, we determine the TA-MSD. Since, the corresponding TA-MSDs for $P_{0, \beta}(\alpha; x, t)$ in the limit of $\tau\to 0$ contain $\langle [x(t + \mathfrak{T}) - x(t)]^2\rangle = \langle x^2(t + \mathfrak{T})\rangle - 2 \langle x(t+\mathfrak{T}) x(t) \rangle + \langle x^2(t)\rangle$, we need to find the autocorrelation function $\langle x(t+\mathfrak{T}) x(t)\rangle$. 

According to Eq.~(4) of Ref.~\cite{AGCherstvy13a} the autocorrelation function for $t_2 > t_1$, which corresponds to the heterogeneous equation in the Stratonovich interpretation, i.e., Eq.~\eqref{23/04/25-1} in the limit of $\tau\to 0$ or Eq.~\eqref{17/12/24-1} for $\alpha = 1/2$, $\lambda = 0$, and $\tau\to 0$, reads
\begin{align}
    \label{10/12/25-1}
    \langle x(t_1) &x(t_2) \rangle_{W_{0, \beta}}  =  \frac{2^\beta}{\sqrt{\pi}} \frac{\Gamma(1+\beta) \Gamma(1+\beta/2)}{\beta^{2\beta} \Gamma([1+\beta]/2)}\, \nonumber \\ 
    & \times \left(B t_1\right)^{(1+\beta)/2}\left[B (t_2 - t_1)\right]^{(\beta - 1)/2}\,\nonumber \\ 
    &\times {_2F_1}\left(
    \begin{array}{c}
    [1-\beta]/2, 1 + \beta/2 \\
    3/2
    \end{array}
    ; -\frac{t_1}{t_2 - t_1} \right).
\end{align}
For the heterogeneous media in which the diffusion coefficient is given by ${\cal D}_{0, \beta}(x)$, the TA-MSD for $\mathfrak{T} \ll T$ can be approximated by Eq.~(21) of Ref.~\cite{AGCherstvy13} and Eq.~(7) of Ref.~\cite{AGCherstvy13a}
\begin{equation}\label{11/12/25-1}
\langle \delta^2(\mathfrak{T}, T) \rangle_{W_{0, \beta}} \sim \frac{\Gamma(1/2 + \beta)}{\sqrt{\pi}} \left(2/\beta\right)^{2\beta} B^{\beta} \frac{\mathfrak{T}}{T^{1-\beta}}.
\end{equation} 
Compared to the ensemble averaged MSD~\eqref{19/01/26-3}, one concludes that 
\begin{equation*}
\langle \delta^2(\mathfrak{T}, T) \rangle_{W_{0, \beta}}=(\mathfrak{T}/T)^{1-\beta}\langle x^2((\mathfrak{T})\rangle, 
\end{equation*}
which means a weak ergodicity breaking, since $\langle \delta^2(\mathfrak{T}, T) \rangle_{W_{0, \beta}}\neq \langle x^2((\mathfrak{T})\rangle$.

\section{Summary}\label{sect6}

In this paper, the diffusion phenomena that emerged in a heterogeneous medium are considered with respect to their possible connections to the noisy voter model. It is a formal but striking similarity of equations governing these models. The medium, which is the arena of heterogeneous diffusion, is characterized by a diffusion coefficient ${\cal D}(x)$. In the studies of stochastic phenomena we have to use different interpretations characterized by the parameter $\alpha\in[0, 1]$. To incorporate this fact to our considerations we completed equations under study adding to them the the additional terms having a meaning of fictitious external forces. In the It\^o form it is equal to ${\cal F}(x) = (1-\alpha) {\cal D}'(x)$. Knowledge of ${\cal D}(x)$ and ${\cal F}(x)$ allows one to find the transition rates $\pi^\pm(x)$ of the voter model. Namely, they read
\begin{equation}\label{27/01/26-1}
    \pi^\pm(x) = \frac{N}{2} {\cal D}(x) \pm \frac{1}{2} {\cal F}(x).
\end{equation}
Here, we considered two types of diffusion coefficients, ${\cal D}_{0, \beta}(x) \propto |x|^{2 - 2/\beta}$ and ${\cal D}_{\lambda, \beta}(x) \propto (\lambda + |x|)^{2 - 2/\beta}$ for $\beta > 0$ and $\lambda \geq 0$. From a physical point of view, the model based on ${\cal D}_{0, \beta}(x)$ has disadvantages related to the fact that at $x = 0$, ${\cal D}_{0, \beta}(x)$ is zero or tends to infinity. Thus, we generalize the diffusion coefficient to ${\cal D}_{\lambda, \beta}(x)$. 

The voter model can be transformed into a heterogeneous Fokker-Planck equation for the PDF, where $x$ is a power of the ratio of the number of voters in the state $1$ to the number of voters in the state $0$, which in financial markets corresponds to the long–term varying component of return. Since the Fokker-Planck equation is a parabolic equation equipped with an infinite propagation speed, in the voter model, this could mean that there is a non-zero probability that some numbers of voters can change their opinion from the initial state $1$ to the state $0$ instantaneously. In reality, such change of decision usually requires time, so we consider a model with a finite propagation speed. Such a candidate is a model governed by the heterogeneous Cattaneo-Vernotte equation. Accepting this approaches, we have proposed to use and solve the Cattaneo-Vernotte equation with ${\cal D}_{\lambda, \beta}(x)$ for various values of the parameter $\alpha \in [0, 1]$. It was done in three cases, namely for $\lambda = 0$ and $\lambda \neq 0$ among them for $\alpha = 1/2$ and $\beta\in\mathbb{R}$, $\alpha = 1$ and $\beta = 3$, as well as $\alpha = 0$ and $\beta = 5$. The solutions are given by symmetric functions for which only the even moments exist. We calculate the corresponding moments and the MSDs. We also show that $P_{\lambda, \beta}(\alpha; x, t)$ in the limit of small $\tau$ and $\alpha = \{0, 1/2, 1\}$ goes to the solution of the corresponding Fokker-Planck equation $W_{\lambda, \beta}(\alpha; x, t)$. We also analyze the TA-MSD and find a weak ergodicity breaking. It can be concluded that the average values of the voting results across the ensemble or over a sufficiently long period of time are the same. However, their variances are completely different and cannot be interchanged.

\section*{Acknowledgment}

KG acknowledges the financial support provided under the NCN Research Grant Preludium Bis 2 No. UMO-2020/39/O/ST2/01563. TP thanks for the financial support under the NAWA Research Grant Preludium Bis 2 No. BPN/PRE/2022/1/00082. TS acknowledges financial support from the German Science Foundation (DFG, Grant number ME~1535/12-1) and from the Alliance of International
Science Organizations (Project No.~ANSO-CR-PP-2022-05). TS was also supported by the Alexander von Humboldt Foundation.

\appendix

\section{Special functions}\label{Ap1}

The three-parameter Mittag-Leffler function is defined by~\cite{TRPrabhakar71}
\begin{equation}
    \label{6/01/26-2}
    E_{a, b}^c(x) = \sum_{r=0}^{\infty} \frac{(c)_r\; x^r}{r! \Gamma(b + a r)}, \quad x\in\mathbb{R},
\end{equation}
where $(c)_r = \Gamma(c + r)/\Gamma(c)$ is the Pochhammer symbol. The parameters $a, b$, and $c$ satisfy the following conditions $a, b, c > 0$ and if $0 < ac < b$ then $E_{a, b}^{c}(\cdot)$ is a completely monotonic function (CMF).~ \cite{KGorska21a}

\begin{definition}
CMF is a function $f: (0, \infty) \mapsto (0, \infty)$ which belongs to ${\cal C}^{\infty}$ and whose derivatives alternate, such that $(-1)^n f^{(n)}(x) > 0$ for $n=0, 1, \ldots$. 
\end{definition}

The Laplace transform of the Prabhakar function $t^{b-1} E_{a, b}^c(-\kappa t^a)$ reads\cite{TRPrabhakar71}
\begin{equation}
    \label{11/01/26-1}
    \mathscr{L}[t^{b-1} E_{a, b}^c(-\kappa t^a); s] = \frac{s^{ac - b}}{(s^a + \kappa)^c}, \quad \kappa\in\mathbb{R},
\end{equation}
For $c = 1$, it reduces to the two-parameter Mittag-Leffler function $E_{a, b}(x)$, and for $b = c = 1$ to the one-parameter Mittag-Leffler function $E_a(x)$.

The one-sided L\'{e}vy stable distribution $\varPhi_{\mu}(\sigma)$ for $\sigma \in (0, 1)$ is nonzero for $\sigma > 0$ and it vanishes for $\sigma \leq 0$. The nonzero value of $\varPhi_\mu(\sigma)$ for $0< \mu = l/k < 1$ can be expressed in terms of the Meijer $G$ function\cite{KAPenson10}
\begin{equation}
    \label{6/01/26-4}
    \varPhi_{l/k}(\sigma) = \frac{\sqrt{kl}}{(2\pi)^{(k-l)/2}} \frac{1}{\sigma} G^{k, 0}_{l, k}\left(\frac{l^l}{k^k \sigma^l}\Big| 
    \begin{array}{c}
    \Delta(l, 0) \\ \Delta(k, 0)
    \end{array}
    \right),
\end{equation}
According to common convention, the special list of $n$ elements is equal to $\Delta(n, a) = a/n, (a+1)/n, \ldots, (a+n-1)/n$. The one-sided L\'{e}vy stable distribution $\varPhi_\beta(\sigma)$ possesses the self-similar property, owing to which we can define the two-variabels L\'{e}vy stable distribution $\varPhi_\mu(\xi, t) \equiv \varPhi_\mu(t \xi^{-1/\mu}) = \xi^{1/\mu} \varPhi_\mu(\xi, t)$. 

\section{Solution $\widehat{P}_{\lambda, \beta}(\alpha; x, s)$}\label{Ap2}

The heterogeneous CV \eqref{17/12/24-1} with diffusion coefficent ${\cal D}_{\lambda, \beta}(x)$ in Laplace space with the fundamental initial conditions reads
\begin{align}\label{28/12/24-1}
    (\tau s^2 + s) & \widehat{P}_{\lambda, \beta}(\alpha; x, s) - (\tau s + 1) \delta(x)  = \frac{B\, \partial_x^2 \widehat{P}_{\lambda, \beta}(\alpha; x, s)}{(\lambda + |x|)^{2/\beta - 2}} \nonumber \\
    & + \frac{2 B\, (\alpha + 1)(\beta-1)}{\beta(\lambda + |x|)^{2/\beta-1}} [2 \Theta(x) - 1] \partial_x \widehat{P}_{\lambda, \beta}(\alpha; x, s)   \nonumber \\
    &  + \frac{B \alpha (\beta - 1)}{\beta (\lambda + |x|)^{2/\beta}} \Big[\frac{\beta - 2}{\beta}  + 2\delta(x)(\lambda + |x|) \Big] \nonumber \\ & \times \widehat{P}_{\lambda, \beta}(\alpha; x, s).
    \end{align}
Setting $X = |x|$, we can express the symmetric function $\widehat{P}_{\lambda, \beta}(\alpha; x, s)$ as $\widehat{G}_{\lambda, \beta}(\alpha; s) \widehat{F}_{\lambda, \beta}(\alpha; X, s)$. Consequently, Eq.~\eqref{28/12/24-1} is transformed into two equations. One is defined for $x \neq 0$,
\begin{align}\label{30/12/24-1}
    &\partial_X^2 \widehat{F}_{\lambda, \beta}(\alpha; X, s) + \frac{2(\alpha + 1)(\beta - 1)}{\beta\, (\lambda + X)} \partial_X \widehat{F}_{\lambda, \beta}(\alpha; X, s) \nonumber\\& + \left[\frac{2\alpha}{\beta^2} \frac{ (\beta - 1) (\beta - 2)}{(\lambda + X)^2} - \frac{\tau s^2 + s}{B (\lambda + X)^{2 - 2/\beta}} \right] \widehat{F}_{\lambda, \beta}(\alpha; X, s) = 0.
\end{align}
while the second one is related to the terms with $\delta(x)$, and it gives the non-zero results only for $x = 0$. Thus, we have
\begin{align}\label{28/12/24-2}
    &-\frac{\tau s + 1}{\widehat{G}_{\lambda, \beta}(\alpha; s)}  = 2 B\Big[(\lambda + X)^{2 - 2/\beta} \partial_X \widehat{F}_{\lambda, \beta}(\alpha; X, s) \nonumber\\  &+  2\alpha (1 - 1/\beta) (\lambda + X)^{1 - 2/\beta} \widehat{F}_{\lambda, \beta}(\alpha; X, s)\Big]_{X = 0}.
\end{align}
The solution of Eq.~\eqref{30/12/24-1} can be obtained by comparing it with the Bessel-type equation 
\begin{equation}\label{20/12/24-1}
    f''(\xi) + \frac{1 - 2 a}{\xi} f'(\xi) + \left(b^2 c^2 \sigma^{2c  -2} + \frac{a^2 - \nu^2 c^2}{\xi^2}\right) f(\xi) = 0,
\end{equation}
with solution $f(\xi) = \xi^a Z_\nu(b \xi^c)$, where $b = \I b_1$,  $Z_\nu(u) = C_1 J_\nu(u) + C_2 Y_\nu(u)$ for integer $\nu$ and $Z_\nu(u) = c_1 J_{\nu}(u) + c_2 J_{-\nu}(u)$ for noninteger $\nu$, see Chapter~104 of Ref.~\cite{FBrowman58}. Here, $C_j$ and $c_j$ for $j =\{1, 2\}$ are the integration constants that can be determined from the boundary conditions.

Setting $\xi = \lambda + X$ in Eq.~\eqref{20/12/24-1}, and requiring that $f(\xi)$ vanishes at $\pm\infty$, we find that $\widehat{F}_{\lambda, \beta}(\alpha; X, s)$ for $\beta > 0$ has the form 
\begin{equation}\label{18/12/24-4}
       \widehat{F}_{\lambda, \beta}(\alpha; X, s) = (\lambda + X)^{a}\, K_{\nu}\left(\frac{\beta\, (\lambda + X)^{1/\beta}}{\sqrt{B}} \sqrt{\tau s^2 + s}\right),
\end{equation}
where $K_{\mu}(\cdot)$ is the modified Bessel function of the second kind (the Macdonald function). In this equation, $a$ and $\nu$ are defined by eqs.~\eqref{5/09/25-1} and \eqref{31/01/26-1}, and $b_1 = \beta\,\sqrt{\tau s^2 + s}/\sqrt{B}$. Substituting it into Eq.~\eqref{28/12/24-2}, for $\beta >0$, we get
\begin{align}\label{28/12/24-3}
    \widehat{G}_{\lambda, \beta}(\alpha; s) =& \frac{\lambda^{(\nu-1)/\beta}}{2 \sqrt{B}} \frac{\tau s + 1}{\sqrt{\tau s^2 + s}} \nonumber\\&\times \left[K_{\nu-1}\!\left(\!\frac{\beta \lambda^{1/\beta}}{\sqrt{B}} \sqrt{\tau s^2 + s}\right)\right]^{-1}.
\end{align}
Multiplying $\widehat{F}_{\lambda, \beta}(\alpha; X, s)$ by $\widehat{G}_{\lambda, \beta}(\alpha; s)$ we find Eq.~\eqref{6/04/25-1}.

\section{Normalization of $P_{\lambda, \beta}(\alpha; x, t)$}\label{Ap3}

The direct calculation shows that $P_{\lambda, \beta}(\alpha; x, t)$ is normalized to $x\in\mathbb{R}$. That is,
\begin{align}
    \label{21/07/25-1}
   & \int_{-\infty}^{\infty} P_{\lambda, \beta}(\alpha; x, t) \D x  = \mathscr{L}^{-1}\left[\int_{-\infty}^{\infty} \widehat{P}_{\lambda, \beta}(\alpha; x, s) \D x; t\right] \nonumber \\ 
    & = \mathscr{L}^{-1}\left[2 \widehat{G}_{\lambda, \beta}(\alpha; s) \int_{0}^{\infty} \widehat{F}_{\lambda, \beta}\big(\alpha; (\lambda + x)^{2-2/\beta}, s\big) \D x; t\right].
\end{align}
For $\beta > 0$ we set $(\lambda + x)^{1/\beta} = \xi$ which allows us to write
\begin{align}\label{21/07/25-2}
   \int_{0}^{\infty} (\lambda + x)^{a} K_{\nu}\big(b_1 (\lambda + x)^{1/\beta}\big) \D x  &= \beta \int_{\lambda^{1/\beta}}^{\infty} \xi^{1-\nu} K_{\nu}(b_1 \xi) \D \xi \nonumber\\
   &= \frac{\beta\lambda^{(1-\nu)/\beta}}{b_1} K_{\nu-1}(b_1 \lambda^{1/\beta}),
\end{align}
where $b_1$ is defined below Eq.~\eqref{18/12/24-4}. Here, we used Eq.~(2.16.3.8) of Ref.~\cite{APPrudnikov-v2}. Its product with  $\widehat{G}_{\lambda, \beta}(\alpha; s)$ gives $\mathscr{L}^{-1}[1/s; t]=1$.

\section{Nonnoegativity of $P_{\lambda, \beta}(\alpha; x, t)$}\label{Ap4}

To prove the nonnegativity of $P_{\lambda, \beta}(\alpha; x, t) = \mathscr{L}^{-1}[\widehat{P}_{\lambda, \beta}(\alpha; x, s); t]$, where $\widehat{P}_{\lambda, \beta}(\alpha; x, s)$ is given by Eq.~\eqref{6/04/25-1}, we will use the Bernstein theorem. The theorem states:
\begin{theorem}
 $f(t)$ is nonnegative if and only if its Laplace transform is a completely monotonic function (CMF) $\widehat{f}(s)$, $s > 0$.
\end{theorem}

We can show that $\widehat{P}_{\lambda, \beta}(\alpha; x, s)$ is a CMF for $\nu\in(0, 3/2]$. The proof is presented in the following and is split into three parts. 

\medskip

\noindent
{\bf Part~1.}~To prove that $\widehat{P}_{\lambda, \beta}(\alpha; x, s)$ is a CMF for $\nu\in(0, 1/2)$, we use that in this range of $\nu$ the function $s^{\nu} \exp(c_{1} s) K_{\nu}(c_{1} s)$, $c_{1} > 0$, is a CMF, see Theorem~5 of Ref.~\cite{KSMiller01}.  As $\exp(- c_{1} s)$ is a CMF we conclude that 
\begin{equation}\label{15/10/25-2}
f_1(s) = s^{\nu} K_\nu(c_{1} s) = \exp(- c_{1} s)\, s^{\nu}\exp(c_{1} s)K_{\nu}(c_{1} s)
\end{equation}
is a CMF. From Remark~1 of Ref.~\cite{KSMiller01}, which states that if $f(s)$ is a CMF then its odd derivatives with the sign minus are also CMF, we have that
\begin{equation}\label{15/10/25-3}
g_1(s) = s^{\nu} K_{\nu-1}(c_{2} s), \quad c_{2} > 0, 
\end{equation}
is also a CMF. Then, according to Corollary~2 of Ref.~\cite{KSMiller01}, if $g$ is a CMF and $g(0) < \infty$, then the function $[D - g(x)]^{-\mu}$, $D\geq g(0)$, $\mu\ge0$ is a CMF as well. Using the asymptotic formula for the Macdonald function $K_{\nu}(s)$ for $\nu >0$ and $s\ll 1$, 
\begin{equation}\label{7/04/25-2}
    K_{\nu}(s) \sim \frac{\Gamma(\nu)}{2} \left(\frac{2}{s}\right)^{\nu}
\end{equation}
we conclude that $g_1(0) = \lim_{s\to 0} s^{\nu} K_{\nu-1}(c_{2} s) = 0$. Taking, for instance, $D = g_1(0)$ and $\mu = 2$, we obtain that $g_1^{-2}(s)$ is a CMF. Hence,
\begin{equation}\label{15/10/25-4}
\frac{f_1(s) g_1(s)}{g_1^{2}(s)} = \frac{f_1(s)}{g_1(s)} = \frac{K_{\nu}(c_{1} s)}{K_{\nu - 1}(c_{2} s)}
\end{equation}
is CMF for $\nu\in(0, 1/2)$.

\medskip
\noindent
{\bf Part~2.}~Due to Eq.~(15.9.126) of Ref.~\cite{RLSchilling10}, one finds that the function ${\sqrt{s}}\exp(c_2 s) K_{\nu-1}(c_2 s)$ for $c_2 >0$ and $\nu\in(1/2, 3/2)$ is a (complete) Bernstein function, i.e., nonnegative function whose first derivative is a CMF, and thus
\begin{equation}\label{15/10/25-5}
    g_2(s) = \frac{1}{\sqrt{s}\,\exp(c_2 s)\, K_{\nu-1}(c_2 s)}
\end{equation}
is a CMF. As indicated in Theorem~5 of Ref.~\cite{KSMiller01}, $f(s) = \sqrt{s} \exp(c_{1} s) K_{\nu}(c_{1} s)$ is a CMF for $\nu > 1/2$ and $c_1 > 0$. Hence,
\begin{align}\label{15/10/25-6}
f_2(s) & = \sqrt{s} \exp(c_{2} s) K_{\nu}(c_{1} s) \nonumber \\ 
& = {\rm e}^{-(c_1 - c_2) s}\, \sqrt{s}\, {\rm e}^{c_{1} s} K_{\nu}(c_{1} s), 
\end{align}
for $c_1 \geq c_2$, is a CMF as well, the same as the product of $f_2(s)$ and $g_2(s)$ for $\nu\in(1/2, 3/2)$. 

\medskip
\noindent
{\bf Part~3.}~The Macdonald functions $K_\nu(s)$ for $\nu = 1/2$ and $\nu = 3/2$ equal to
\begin{multline}\label{7/04/25-1}
K_{1/2}(s) = \sqrt{\frac{\pi}{2 s}}\, \exp(-s) \quad \text{and} \\ K_{3/2}(z) = \sqrt{\frac{\pi}{2 s}}\, \exp(-s) \Big(1 + \frac{1}{s}\Big).
\end{multline}
Hence,
\begin{multline}
    \label{15/10/25-7}
    \frac{K_{1/2}(c_1 s)}{K_{-1/2}(c_2 s)} = \sqrt{\frac{c_2}{c_1}}\, {\rm e}^{-(c_1 - c_2) s} \quad \text{and} \\ \frac{K_{3/2}(c_1 s)}{K_{1/2}(c_2 s)} = \sqrt{\frac{c_2}{c_1}} {\rm e}^{-(c_1 - c_2) s}\,\left(1 + \frac{1}{c_1 s}\right).
\end{multline}
Both ratios of the Mcdonalds functions are CMFs for $c_1 \geq c_2$. \qed

\section{Inverse Laplace transform of Macdonald function}\label{Ap5}

Expressing Eq.~\eqref{9/09/25-1} in the form
\begin{align}
    \label{19/04/25-2}
\frac{s + \kappa + \kappa}{[(s + \kappa)^2 - \kappa^2]^{\nu/2}} K_\nu\left(C \sqrt{(s + \kappa)^2 - \kappa^2} \right),
\end{align}
where $\kappa = 1/(2\tau)$ and $C = \alpha |x|^{1/\alpha}/\upsilon$, $\upsilon = \sqrt{B/\tau}$, we calculate its inverse Laplace transform using the well-known property $\mathscr{L}^{-1}[\hat{f}(s + \kappa); t] = {\rm e}^{- \kappa t} \mathscr{L}^{-1}[\hat{f}(s); t]$. Thus, we get
\begin{equation}
    \label{19/04/25-3}
    {\rm e}^{-t \kappa} \mathscr{L}^{-1}\left[\frac{s + \kappa}{(s^2 - \kappa^2)^{\nu/2}} K_\nu\Big(C \sqrt{s^2 - \kappa^2}\Big); t \right].
\end{equation}
Eq.~\eqref{19/04/25-3} is calculated employing Eq.~(3.16.2.2) from Ref.~\cite{APPrudnikov-v5}, 
\begin{multline}
    \label{19/04/25-4}
   \mathscr{L}^{-1}[(s^2 - \kappa^2)^{-\nu/2} K_\nu(C \sqrt{s^2 - \kappa^2}); t] \\ = \sqrt{\frac{\pi}{2}} \frac{\kappa^{1/2-\nu}}{C^\nu}  (t^2 - C^2)_+^{\frac{\nu}{2} - \frac{1}{4}} I_{\nu - \frac{1}{2}}(\kappa\sqrt{t^2 - C^2})
\end{multline}
and we also apply the formula of the Laplace transform of derivative, $\mathscr{L}^{-1}[s\hat{f}(s); t] = \frac{\rm d}{{\rm d} t}f(t) + f(0)$. Note that Eq.~\eqref{19/04/25-4} at $t=0$ is equal to zero. Hence, 
\begin{align}
    \label{19/04/25-5}
    &\mathscr{L}^{-1}[s(s^2 - \kappa^2)^{-\nu/2} K_\nu(C \sqrt{s^2 - \kappa^2}); t]\nonumber\\& = \sqrt{\frac{\pi}{2}} \frac{\kappa^{1/2-\nu}}{C^\nu} \left\{\!\delta(t - C)+ \kappa t (t^2 - C^2)_+^{\frac{\nu}{2} - \frac{3}{4}}  I_{\nu - \frac{3}{2}}(\kappa \sqrt{t^2 - C^2})\!\right\}. 
\end{align}
The symbol $u_{+}^{\lambda}$ is equal to $u^{\lambda}$ for $u > 0$ and $0$ for $u < 0$. Note that Eqs.~\eqref{19/04/25-4} and~\eqref{19/04/25-5} for $\nu = 1/2$ reduce to the known result reported in Appendix A of Ref.~\cite{JMasoliver19} and in Appendix~B of Ref.~\cite{YPovstenko21}.

We also use formula~(80) from Appendix~B of Ref.~\cite{YPovstenko21}, which for $\Lambda > 0$ reads
\begin{align}\label{12/09/25-6}
&\mathscr{L}^{-1}\big[\exp(-\Lambda \sqrt{s^{2} - \kappa^{2}}); t\big] \nonumber\\&= \delta(t - \Lambda) + \Theta(t-\Lambda) \frac{\Lambda \kappa}{\sqrt{t^{2} - \Lambda^{2}}} I_{1}(\kappa \sqrt{t^{2} - \Lambda^{2}}).
\end{align}

\section*{References}

\nocite{*} 
 \bibliography{Refs1}
\end{document}